\documentclass[final]{aipproc}

\usepackage{graphicx}

\layoutstyle{6x9}

\usepackage{epsfig}

\begin{document}

\title{Physical problems (microphysics) in relativistic plasma flows} 

\classification{}
 \keywords{plasma instabilities; numerical simulations; B-field generation; particle acceleration}

\author{L. O. Silva}{
  address={GoLP/Centro de F\'isica dos Plasmas, Instituto Superior T\'ecnico, 1049-001 Lisboa, Portugal}
}

\begin{abstract}
 Many problems in astrophysics involve relativistic outflows. The plasma dynamics in these scenarios is critical to determine the conditions for the self-consistent evolution of the fields and particle acceleration. Advances in computer power now allow for kinetic plasma simulations, based on the particle-in-cell (PIC) paradigm, capable of providing information about the role of plasma instabilities in relativistic outflows. A discussion of the key issues associated with PIC simulations is presented, along with some the most important results and open questions, with a particular emphasis on the long time evolution of the filamentation, or Weibel, instability, and on the possible collisionless mechanisms for particle acceleration arising in the collision of relativistic plasma shells.
\end{abstract}

\maketitle


\section{Relativistic outflows and gamma-ray bursts}

Relativistic outflows are pervasive in astrophysics, from jets to gamma-ray bursts, from accretion disks to supernovae explosions. Some of the underlying physical mechanisms in these flows have strong connections with the dynamics and the physics of plasmas, and many outstanding problems in astrophysics are closely associated with scenarios where the collisionless dynamics of plasmas can play an important role. The obvious examples are the formation of relativistic shocks (and particle acceleration in these structures), magnetic field generation due to kinetic plasma instabilities, or non-thermal particle acceleration.

The onset of these processes is associated with collisionless plasma instabilities. There is a wealth of theoretical work on these instabilities, going back to the early days of plasma physics, but only now, with the advent of massivelly parallel computing, it is possible to perform realistic detailed numerical simulations of these instabilities, in order to understand not only the linear, transient, stage of these instabilities but also the long time saturated behavior of the scenarios where such instabilities can occur. This is opening the way to establish connections between the plasma dynamics at the kinetic level and its consequences in different astrophysical phenomena.

A clear example of the importance of plasma instabilities in relativistic ouflows is associated with the fireball model of gamma rays bursts (GRBs) (cf. R. Sari, this conference \cite{sari2006}, and \cite{piran2004}, and references therein). In the fireball model of GRBs \cite{meszaros1993}, relativistic plasma shells collide/overtake each other, leading to the rapid variability of the observed radiation. The radiation is believed to be from synchrotron origin, which requires sub-equipartition magnetic fields to be generated in the collision of the plasma shells, and to survive for time scales much longer than the collisionless time scales. One possible mechanism that can explain the generation of magnetic fields at these levels in GRBs is the Weibel instability \cite{medvedev1999,gruzinov1999}. Recently, numerical simulations have strengthened this conclusion  \cite{silva2003,frederiksen2004,jaroscheck2004,hededal2005,nishikawa2005}.   

In this particular problem, many questions remain to be fully addressed, whose answers are general enough to be of interest to other problems in astrophysics. What is the long time evolution of the magnetic fields generated via collisionless instabilities? What are the consequences of the field structure generated by collisionless instabilities to the radiation observed from these objects? How are particles accelerated in the fields resulting from the collision of relativistic plasma flows? Can the fields lead to the formation of relativistic shocks? How are particles accelerated in relativistic shocks? While some answers have already been proposed,  kinetic simulations, regarded as a numerical laboratory for astrophysics, combined with relativistic kinetic theory, can lead to significant progress in solving some of these open questions.

In this paper, I will review the basic concepts behind kinetic plasma simulations, and their most common paradigm, pointing out some of the advantages and some of the difficulties of this technique. Relevance will be given to the possibility to probe the microphysics on the time scale of the electron collective dynamics, and to obtatin detailed information about the structure of the fields and the distribution function of the particles. In Section III, the most recent developments on collisionless instabilities in unmagnetized plasmas will be described, illustrated with numerical simulations. The limits posed to particle-in-cell simulations of collisionless instabilities are discussed, and different strategies to overtake these limits are proposed. In Sections IV and V, the exploration of magnetic field generation and particle acceleration using lower dimensionality simulations is presented, employing alternative physical configurations. Generation of sub-equipartition magnetic fields is confirmed, along with the two step evolution of the filamentation instability (on the electron time scale, and on the ion time scale), in simulations running for times three orders of magnitude longer than the electron dynamics time scale. Particle acceleration is also observed in scenarios resulting from the collision of relativistic plasma shells, associated with (i) the filament coalescing process \cite{silva2003}, (ii) acceleration in the electric field of the filaments generated in the Weibel instability \cite{hededal2004}, and (iii) acceleration in the relativistic electron plasma wave generated in the interface region between the two colliding shells \cite{silva2006}. Finally, in Section VI, I state the conclusions and I point out the open question to which PIC simulations might contribute in the near future.

\section{Particle-in-cell simulations}

Particle-in-cell (PIC) simulations are one of the most common numerical tools in plasma physics \cite{dawson1983,birdsall1985}. Originated in the pioneering work of Oscar Buneman and John Dawson in the 1960s, it has evolved to a mature technique commonly used in many sub-fields of plasma physics. 

\begin{figure}[hbt]
\epsfig{file=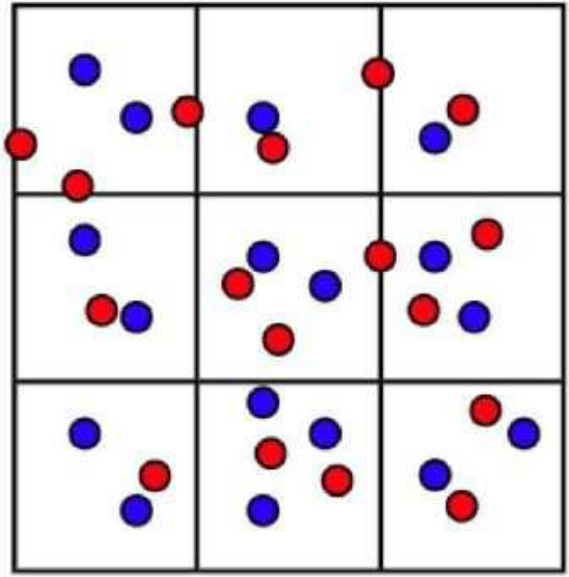,width=8 cm}
\caption{General scheme of PIC simulations: the electric field and the magnetic field are calculated in two staggered grids, while charged particles move in all regions of space.}
\label{fig:1}
\end{figure}

The idea behind particle-in-cell simulations is quite simple. The motion of a set of charged particles is followed under the action of the self-consistent fields generated by the charged particles themselves. Maxwell's equations are solved on a grid, with the sources for the equations for the field advance (current and/or charge) determined by depositing the relevant quantities from the particles on the grid (Figure \ref{fig:1}). 

\begin{figure}[hbt]
\epsfig{file=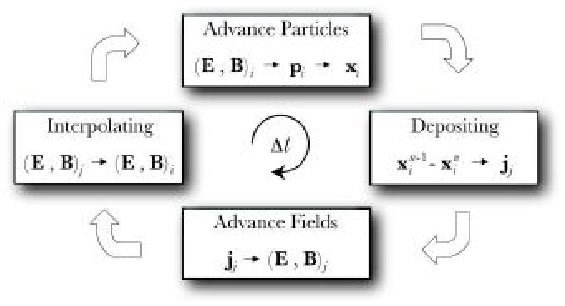 ,width=8 cm}
\caption{General scheme of a PIC loop \cite{fonseca2002c}.}
\label{fig:1b}
\end{figure}

After advancing the fields in time, the information to determine the fields on each particle's position is available, the Lorentz force on the charged particles can be determined, the particles can be pushed, and their position and momentum updated to the new values. After the particle advance, the new quantities to advance the fields can be calculated, thus closing the loop (Figure \ref{fig:1b}).  

To the extent that quantum mechanical effects can be neglected, these codes make no physics approximations and are ideally suited for studying complex systems with many degrees of freedom. The advent of massivelly parallel computing now allows for PIC simulations using more than $10^9$ particles, in systems with $(500)^3$ cells, using 0.5 TByte of RAM, with runtimes from a few hours to several weeks in computing systems with 100s of CPUs, producing data sets that can easily reach the TByte level. For some problems in plasma based accelerators \cite{tsung2006}, three-dimensional PIC one-to-one simulations with the exact experimental parameters are already performed, complementing the experimental diagnostics, and acting as virtual experiments.  

The information available from PIC simulations provides an outstanding tool to test new models and new ideas, but the complexity of developing, maintaining, and running these codes, and exploring the data generated in these simulations, requires research teams with diverse skills, and with dimension comparable to those running small-medium scale experiments. 

The possibility to include new physics has also been explored, and it is now quite common to find massivelly parallel PIC codes, such as OSIRIS \cite{fonseca2002b}, that include impact and tunnel ionization, binary collisions, radiation damping, and that can provide, on post-processing, information about the radiation spectra \cite{hedekal2005b}. Other authors are also attempting to generalize the PIC technique to non flat metrics (Watson {\it et al} \cite{watson2006}, and \cite{daniel1997}), in order to employ PIC codes to model conditions with strong gravitational fields, for instance in the vicinity of black holes. 
In the astrophysics of relativistic flows, only very recently 3D numerical PIC simulations have become more common \cite{silva2003,frederiksen2004,jaroscheck2004,hededal2005,nishikawa2005}, even though pioneering work in lower dimensional PIC simulations relevant for relativistic astrophysics was undertaken in the 80s (e.g. \cite{leboeuf1982,langdon1988}. However, and for many years, PIC simulations have been critical to understand the nonlinear evolution of many collisionless plasma processes.

\section{Collisionless plasma instabilities in unmagnetized plasmas}

The general theory for collisionless plasma instabilities was developed in the 1960s by Watson, Bludman and Rosenbluth \cite{watson1960}. For a review of the different collisionless instabilities see \cite{davidson}. 
In a unmagnetized plasma, and on the electron collective dynamics time scale, the key instability that can operate when different plasma streams collide is the electromagnetic beam-plasma instability. A detailed and global analysis of the electromagnetic beam-plasma instability was undertaken only very recently. Usually, the limiting scenarios described in the litterature deal only with coupling with the longitudinal electrostatic mode (two stream instability) i.e. the collective electrostatic mode (electron plasma wave) whose wave vector is parallel to the direction of propagation of the plasma stream providing the free energy to the instability, or coupling with the purely transverse electromagnetic mode (Weibel instability) \cite{weibel1959}, with wave vector transverse to the plasma stream propagation direction.

For the two-stream instability, and in the most favorable conditions, the growth rate of the instability $\Gamma$ scales with the ratio between the electron density in the two streams $\Gamma \propto \left(n_b/n_0\right)^{1/3}$, where $n_b$ is the electron density of the (lower density) stream, and $n_0$ is the electron density of the background stationary plasma. The wave number for maximum growth is $k_\| \simeq \omega_{pe0}/v_b$, where $\omega_{pe0}= \left(4 \pi e^2 n_0/m_e\right)^{1/2}$ is the electron plasma frequency and $v_b$ is the velocity of the plasma stream, with $e$ the electron charge, and $m_e$ the electron mass. 

The Weibel instability grows with $\Gamma \propto \left(n_b/n_0\right)^{1/2}$, with a typical wave number $k_\perp \simeq  \omega_{pe0}/c$, for a warm plasma stream, where $c$ is the velocity of light in vacuum. The typical length scales and time scales are determined solely from the electron density, and given by
\begin{equation}
\lambda_e = \frac{c}{\omega_{pe0}} \simeq \frac{5\, \mathrm{km}}{\sqrt{n_0 [ \mathrm{cm^{-3}}]}}
\end{equation}
\begin{equation}
\tau_e = \frac{1}{\omega_{pe0}} \simeq \frac{20\, \mu \mathrm{s}}{\sqrt{n_0 [ \mathrm{cm^{-3}}]}}
\end{equation}

\begin{figure}[hbt]
\epsfig{file=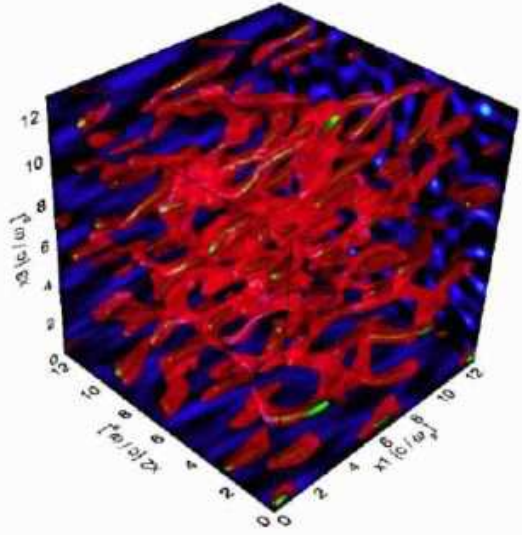,width=8 cm}
\caption{Isosurfaces of the electron density of a relativistic stream undergoing the electromagnetic beam-plasma instability, in the early stages of the instability ($t \simeq 7 \Gamma \tau_e$)}
\label{fig:2}
\end{figure}

In general, however, the collision of plasma streams will drive a combination of the two-stream and the Weibel instability, which should be more accurately described as the electromagnetic beam-plasma instability \cite{silva2001,silva2002,bret2005}. Figure \ref{fig:2} illustrates the typical scenario, arising from the collision of a relativistic plasma stream ($n_b=0.1$, $\gamma_b = 1/\sqrt{1-v_b^2/c^2}=10$ along the $x_1$ direction, with a proper thermal spread $u_{th}=\gamma_b v_{th}=0.1$) with a stationary background plasma ($n_0=1$). In all the simulations presented here, normalized units are employed such that time is normalized to $\tau_e$, lengths are normalized to $\lambda_e$, mass to $m_e$, and charge to $e$.  

\begin{figure}[hbt]
\epsfig{file=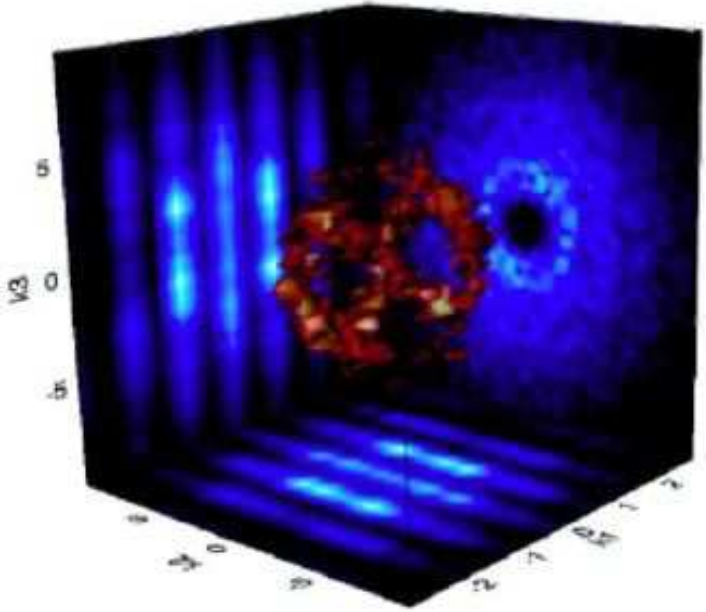,width=4.0 in}
\caption{Distribution of the Fourier modes of the electron density in Figure \ref{fig:2}, showing the tilted nature of the filaments.}
\label{fig:3}
\end{figure}

Strong filamentation of the plasma stream is observed, but the filaments show some degree of tilting, which clearly indicates coupling between the Weibel instability and the two-stream instability. The filament tilting is best shown in the Fourier transform of the electron density (Figure \ref{fig:3}). 
In Figure \ref{fig:3}, the typical wavenumber along the propagation direction is, in our normalized units, $k_1 = k_\| \simeq 1$, and the transverse wavenumber $k_\perp \simeq 3$, which yields a typical angle for the tilted filaments $\theta_\mathrm{tilt} \simeq 18.5^\circ$. 

\begin{figure}[hbt]
\epsfig{file=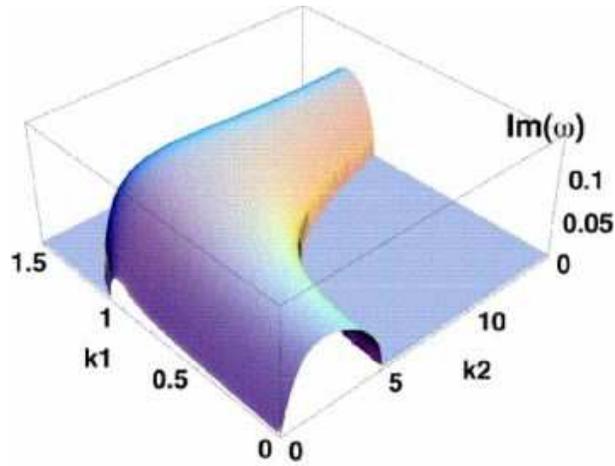,width=8 cm}
\caption{Growth rate of the electromagnetic beam-plasma instability as a function of the wavenumber in two directions ($k_1$ is along the electron stream propagation (longitudinal) direction, and $k_2$ is along the transverse direction), for the conditions of Figure \ref{fig:2}}
\label{fig:4}
\end{figure}

By solving the dispersion relation of the electromagnetic beam-plasma instability for the growth rate of the electromagnetic beam plasma instability it is possible to identify not only the origin of the filament tilting, but also the limiting cases of the two-stream instability and the Weibel instability. As shown in Figure \ref{fig:4}, along the propagation direction ($k_2=0$) the growth rate of the electromagentic beam-plasma instability reduces to the behavior of the two-stream instability, while in the purely transverse direction ($k_1=0$) the typical growth rate for the Weibel is recovered. However, the maximum growth rate occurs for a mode which is slightly off-axis, resulting from the combination of the most unstable wavenumber of the two-stream instability with the most unstable wavenumber of the Weibel instability. 

\begin{figure}[hbt]
\epsfig{file=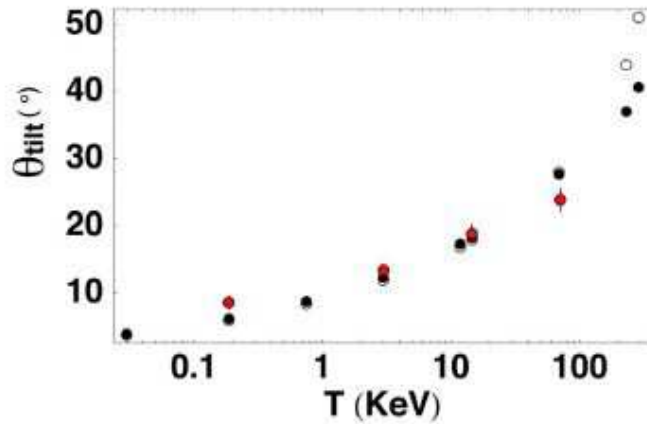,width=3.5 in}
\caption{Angle of tilting of the filaments in the electromagnetic beam-plasma instability as a function of the temperature of the relativistic stream, as measured in two dimensional simulations (black or white circles), and as predicted from the theory of the electromagnetic beam-plasma instability (red/gray circles, with error bar). Same conditions as in Figure \ref{fig:2}, except for transverse stream temperature}
\label{fig:5}
\end{figure}

The angle of tilting depends on the temperature of the relativistic stream. This is natural, since the wavenumber for maximum growth of the Weibel instability has a strong dependence on the temperature \cite{davidson, silva2002}. The angle of tilting can be determined by solving the dispersion relation for the electromagnetic beam-plasma instability \cite{silva2001,bret2005}, and compared with the simulation results, Figure \ref{fig:5}.

The presence of a tilting angle in the filaments is critical for the long time evolution of the filaments. When the plasma stream is warm, or if the instability itself leads to heating of the downstream electrons, the filaments will merge in a rather complex way, preventing the existence of well-formed filaments in regions too far downstream ($\simeq$ 100s $\lambda_e$) from the collision region. This merging will occur on time scales much shorter than the time scales required for long wavelength filament instabilities to occur (e.g. the kink instability \cite{milosavljevic2005}, or the hosing instability \cite{mori2001}). On the other hand, for very cold plasma streams, the instability will be mostly of a two-dimensional nature \cite{leelampe,medvedev2005}.

Another important aspect of the filamentation instability that was overlooked until very recently was the influence of space charge effects. As the electron filaments start to form, an ion channel is present \cite{leelampe}. The corresponding electric field, which tries to prevent the formation of the filaments, should be taken into account in order to obtain the correct thresholds for the occurrence of the filamentation instability \cite{tzoufras2006}. Furthermore, a theoretical analysis of the thresholds and growth rates for the filamentation instability, including space charge effects as well as warm plasma streams with arbitrary mixture of electrons/protons/positrons, showed that the presence of baryons increases the robustness of the instability, increasing the range of unstable temperatures (and wavenumbers) \cite{fiore2006}.

\begin{figure}[hbt]
\epsfig{file=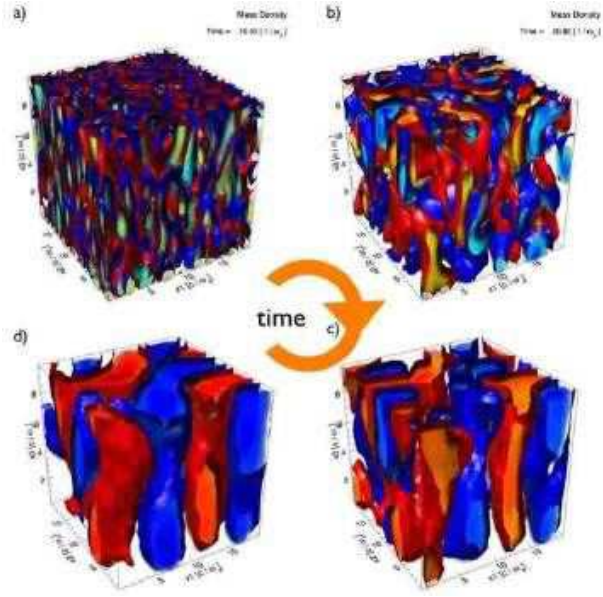,width=8 cm}
\caption{Temporal evolution of the mass density in the filamentation instability. The plasma shells move in the vertical direction}
\label{fig:6}
\end{figure}

\begin{figure}[hbt]
\epsfig{file=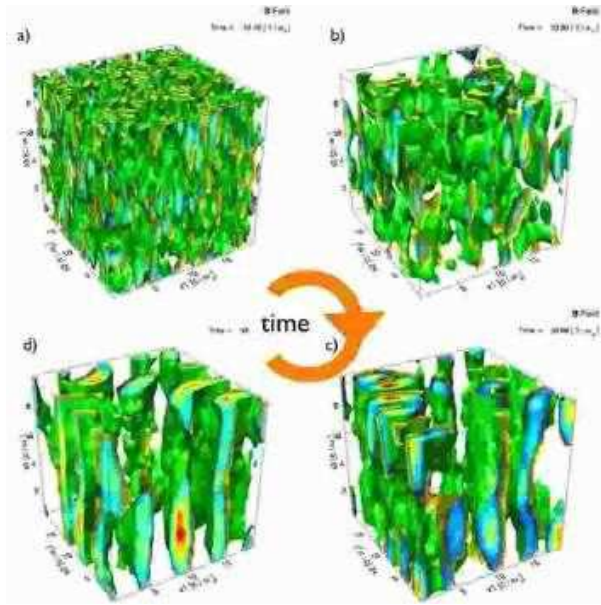,width=8 cm}
\caption{Temporal evolution of the energy in the magnetic field}
\label{fig:7}
\end{figure}

Three-dimensional PIC simulations, performed by many groups around the World \cite{silva2003, frederiksen2004, jaroscheck2004, hededal2005, nishikawa2005, spitkovsky2006, fonseca2003, fonseca2002a}, have confirmed the role played by the electromagnetic beam-plasma instability in scenarios with colliding relativistic plasma shells. The typical temporal evolution is shown in Figures (\ref{fig:6},\ref{fig:7}), obtained for the simulation parameters of \cite{silva2003,fonseca2003}, describing the collision of two relativistic electron-positron plasma shells. The instability evolves from small scale filaments to large scale filaments, as illustrated by the isosurfaces of the mass density of Figure \ref{fig:6}, where blue represents the mass density with positive current density (in the vertical direction), and red represents negative current density. Merging of the filaments is accompanied by the slowdown of the shells, and strong heating \cite{silva2003}. The magnetic field separates regions with opposite current density (cf. Figure \ref{fig:7}), and after saturation (a few 100s $\tau_e$) the magnetic field achieves sub-equipartition levels, with the ratio between the energy density in the B-field and the initial kinetic energy density on the order of $~ \, 10^{-3}-10^{-5}$. Such scenario has been verified in the relativistic case \cite{silva2003}, and in non-relativistic scenarios \cite{medvedev2006a}.

The influence of the magnetic field structure, and its temporal evolution, on the radiation spectra was first pointed out by Medvedev \cite{medvedev2000} (see also \cite{medvedevconf,fleishmanconf}), which demonstrated that the short scale structure of the B-field perturbations due to the Weibel instability would lead to distinct spectral and polarization features. The predictions of the theoretical calculations were recently confirmed with synthetic spectra determined directly from PIC simulations \cite{hedekal2005b}. 

Three-dimensional PIC simulations have indicated the viability of the Weibel instability as the source of sub-equipartition B-fields for times on the order of 100s T. However, the simulations were performed for short times (compared with typical relevant times for gamma ray bursts), and low mass ratios (either electron-positron plasmas, or electron-"proton" plasmas with mass ratios $~10$). 
The long-time evolution has only been addressed with two-dimensional simulations of cold shells, in the plane perpendicular to the bulk velocity direction. For these conditions, a theoretical model can be built, predicting a long time evolution of the filaments that confirms the presence of sub-equipartition B-fields, even for time/length scales much longer than the typical collisionless time/length scales \cite{medvedev2005}. 

\begin{figure}[hbt]
\epsfig{file=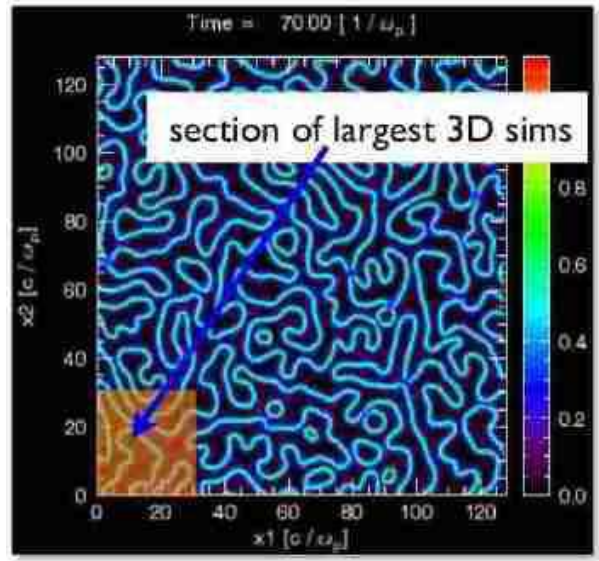,width=8 cm}
\caption{Comparison between the typical length scales covered by three-dimensional and two-dimensional particle-in-cell simulations}
\label{fig:8}
\end{figure}

The difficulty to perform three-dimensional simulations is obvious, given the computational requirements for these large scale simulations. As can be seen in Figure \ref{fig:8}, a medium scale 2D simulation can easily cover length scales that can  encompass even the heavier species length scale. However, and given the possibility that two-dimensional simulations can be performed either in a plane perpendicular to the bulk velocity of the streams, or in a plane perpendicular to the fluid velocity of the streams, these reduced simulations must be performed with care.

Given the length and time scales that can be probed with very large scale two-dimensional simulations, the possibility to push our understanding of astrophysical scenarios with two-dimensional simulations should not be discarded. In order to take advantage of lower dimensionality simulations, it is important to assess how well the relevant physics can be captured. 

\begin{figure}[hbt]
\epsfig{file=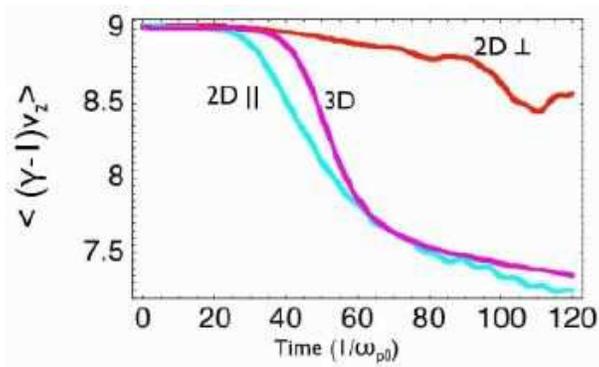,width=8 cm}
\caption{Heat flux along the relativistic stream propagation direction: comparison between 3D and reduced two-dimensional configurations (2D$\|$, and 2D$\perp$) }
\label{fig:9}
\end{figure}

\begin{figure}[hbt]
\epsfig{file=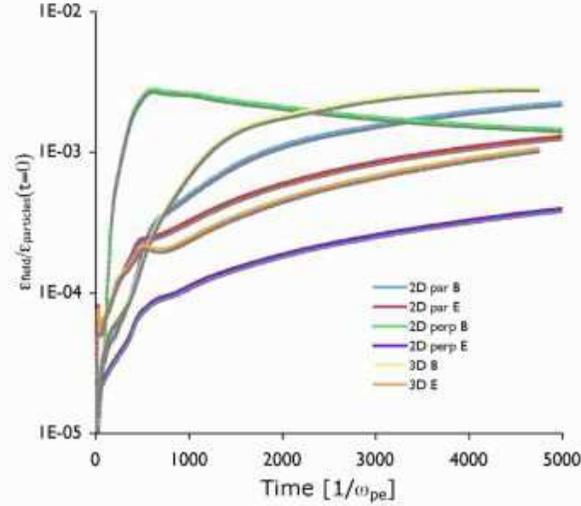,width=8 cm}
\caption{Temporal evolution of the energy in the electric field and the energy in the magnetic field in 3D, and reduced 2D simulations, for conditions leading to the occurrence of the Weibel instability}
\label{fig:10}
\end{figure}

Examples of studies performed to assess the validity of 2D simulations to understand scenarios where the electromagnetic beam-plasma instability occurs are given in Figures (\ref{fig:9}, \ref{fig:10}).    

The Weibel instability can be captured in two-dimensional simulations, either performed in a plane parallel to the velocity of the relativistic shells (2D$\|$), or in simulations in the plane perpendicular to the fluid velocity of the plasma shells (2D$\perp$). In Figure \ref{fig:9}, the heat flux in the forward direction is compared for the different configurations employing identical physical and numerical parameters (relativistic beam with $\gamma_b = 10$, $n_b=0.1 n_0$, streaming through stationary plasma: 2D$\|$ simulations seem to capture more accurately the relevant behavior. 

The same conclusions are obtained when we examine the temporal evolution of the energy in the fields either in the relativistic or the non-relativistic case (cf. Figure \ref{fig:10}, obtained for the conditions of \cite{medvedev2006a} -- non-relativistic shells, with ion to electron mass ratio of 100). Physically the improved agreement between 2D$\|$ and 3D simulations of these scenarios arises from the possibility to excite modes with the longitudinal component of the electromagentic beam-plasma instability. For scenarios where the shells have comparable densities and the velocities are relativistic, coupling with the longitudinal component is important and needs to be taken into account as can be hinted from the comparable values of the two-stream and the Weibel instability in these conditions. In 2D$\|$ simulations this is naturally included, but it is lacking in 2D$\perp$ simulations, thus making 2D$\perp$ simulations more relevant for scenarios with cold shells where coupling with the transverse (filamentation) mode is dominant.

With these results in mind, very long large scale two-dimensional simulations in the plane parallel to the velocity of the relativistic streams (2D$\|$) have been performed in order to examine the long time evolution of the electromagnetic beam-plasma instability. 
Furthermore, since a realistic mass ratio between protons and electrons leads to still too demanding simulations (in particular, if we intend to run very large simulation boxes for very long times), a lower mass ratio between "protons" and ions was used ($m_{"p"}/m_e = 100$). We observe that only mass ratios in this range allow for a clear separation of the electron collective dynamics from the "proton" collective dynamics. The key results from these simulations will be outlined in the next sections. 

\section{Sub-equipartition B-field generation}

The question of whether the Weibel, or filamentation, instability can be the source of sub-equipartition of magnetic fields or not in scenarios relevant for gamma ray bursts has lead to a large number of theoretical and numerical work, as discussed in the previous sections. It is now clear that in pure electron-positron shells the generated B-field reaches sub-equipartition levels. However, for scenarios with electrons and protons, simulations are quite difficult since it is necessary to resolve both the electron length/time scales ($\lambda_e$ and $\tau_e$), and the ion scales $\lambda_i \simeq \left(m_i/m_e\right)^{1/2} \lambda_e$, and $\tau_i \simeq  \left(m_i/m_e\right)^{1/2} \tau_e$. 
So far, simulations have only addressed scenarios with small mass ratios, incompatible with a clear separation 
of the electron scales and ion scales.

Using the  possibility of performing very large 2D$\|$ simulations  more realistic scenarios can be probed. These large simulations also allow for a more detailed analysis of the possible configurations. For 2D$\|$ simulations, different physical configurations can be set-up: (I) a plasma shell, with a non-zero fluid velocity, can be launched continuously from one of the box boundaries (cathode) streaming to a stationary background plasma,  (II) the simulation box moves at the speed of light, and a plasma shell moving with a relativistic fluid velocity is initialized inside the simulation box, while fresh stationary plasma is sent into the simulation box from the boundary where the simulation window is moving to, or (III) two plasma shells, with opposite fluid velocities can be initialized, one of the shells filling the left-hand side of the simulation box, while the second shell fills the right-hand side of the box \cite{gruzinov1999,kazimura}. 

The first scenario (I) describes the interaction of the ejecta with the stationary plasma near the injection region and, if the simulation box is long enough, it can probe the formation of the forward shock and the downstream region of the forward shock. This simulation describes a realistic scenario until the particles of the relativistic plasma shell start to leave the simulation box in the boundary opposite to the injection boundary. It also gives a realistic description of the region near the plasma injection if we assume that the influence of the leading edge of the relativistic shell can be neglected near the injection region.

In the second case (II), the simulation follows the leading edge of the relativistic plasma shell; it allows for the study of the dynamics at the discontinuity, the shock precursor physics and, if the box is long enough, the formation of the relativistic forward shock. It must be stressed that in this scenario all quantities are determined in the reference frame of the stationary (background) plasma even though the simulation box moves at the speed of light. Due to the fact that the window moves at the speed of light it is also clear that whatever conditions are left behind the simulation box, no causal relation can be established between the physics outside the simulation box and the physics observed inside the simulation box (information outside the simulation box is lost).

The third scenario (III) describes the physics in the the center-of-mass reference frame of the two plasma shells. Here it is possible to follow both the contact discontinuities, and the eventual formation of the forward shock and the reverse shock.

\begin{figure}[hbt]
\epsfig{file=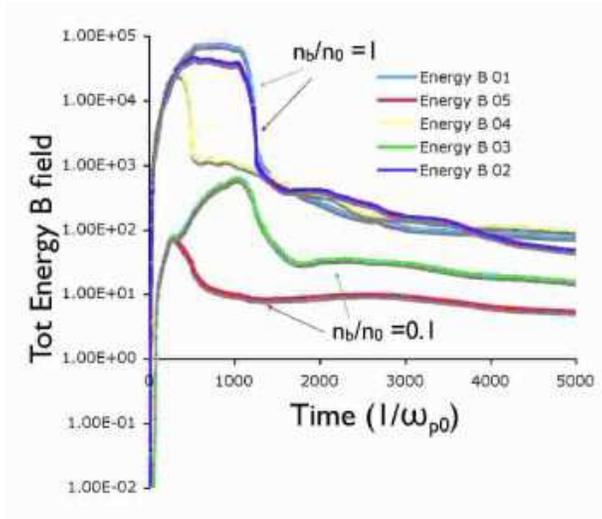,width=8 cm}
\caption{Temporal evolution of the magnetic field for different simulation conditions, involving injection of a plasma stream from the left-hand boundary of the simulation box. In simulations 01, 02, 04 the ratio of the injected plasma density to the stationary plasma density $n_b/n_0$ is 1, while in the simulations 03 and 05, $n_b/n_0=0.1$. In the simulations 01, 02, and 03 the plasma is injected from $t=0$ to $t=1000 \, \tau_e$, while in the simulations 04 and 05 plasma is injected from $t=0$ up to $t=256 \, \tau_e$. $\gamma$ of the injected plasma is 5 for simulation 02, and 10 for all other simulations. Both the injected plasma and the stationary plasma are composed of electron and "protons" with a mass ratio of $m_p/m_e=100$. The simulation box is $(256)^2 \, \lambda_e^2$, $(1280)^2$ cells, with 16 particles/(cell species).}
\label{fig:11}
\end{figure}

In this paper, we will consider scenarios I and II. The evolution of the magnetic field generated in cathode-like configurations (I) is shown in Figure \ref{fig:11}, where the temporal evolution of the total magnetic field inside the simulation box is shown for different physical conditions.

While the cathode in on, the level of the magnetic field is high, at peak saturation levels, sustained by the continuous injection of energy into the simulation region. As the cathode is turned off, the B-field decays to sub-equipartition levels, then slowly decaying on a time scale much longer than the electron or "proton" time scale. Since the peak magnetic field is proportional to the growth rate of the instability \cite{yoon1987,yang1994}, simulations with lower density ratios reveal lower total magnetic fields (even when corrected by the fact that the injected energy density is also lower by a factor of $n_b/n_0$).

\begin{figure}[hbt]
\epsfig{file=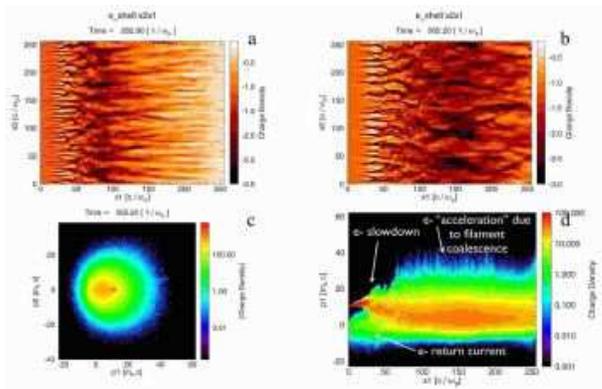,width=8 cm}
\caption{The Weibel instability in a cathode-like configuration (simulation 01): (a) and (b) electron density, (c) $p_1 p_2$ phase space, and (d) $p_1 x_1$ phase space, at late times (corresponding to the same time of (b)). Plasma is injected from the left boundary of the simulation box (in (a) and (b).}
\label{fig:12}
\end{figure}

The temporal evolution of the electron density of the injected relativistic plasma in the configuration I is shown in Figure \ref{fig:12}(a,b), with the region of the electron Weibel instability in (a), from $x_1=0$ to $x_1 \simeq 20 \lambda_e$, followed by the region where the filaments interact and merge (from $x_1 \simeq 20 \lambda_e$ to $x_1 \simeq 50 \lambda_e$. At later times, the ion Weibel instability is also already clear $x_1 \simeq 40 \lambda_e$, seeded by the electron Weibel instability, with similar evolution but on much longer transverse length scales. Figure \ref{fig:12}(c) illustrates the strong isotropic thermal spread affecting the plasma shell (whose $p_1 p_2$ phase space plot at injection reduces to a small region around the red region in \ref{fig:12}(c)). Such isotropization is due to the nonlinear evolution and merging of the filaments. The $p_1 x_1$ phase space, Figure \ref{fig:12}(d),  illustrates the most interesting aspects of the longitudinal electron dynamics in the plasma shell, showing the shell slowdown, and the electron acceleration and strong heating in the region of filament coalescence (to be discussed in the next section). Furthermore, an electron return current is also present, which further enhances the occurrence of the Weibel instability \cite{ren2004}.

\begin{figure}[hbt]
\epsfig{file=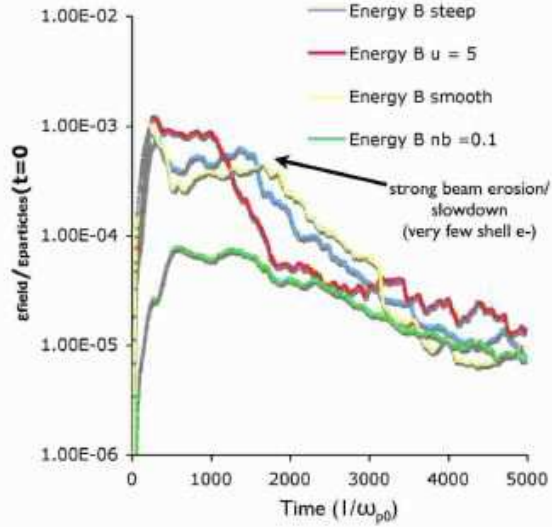,width=8 cm}
\caption{Temporal evolution of the magnetic field for different simulation conditions, in a moving window simulation box (simulation box moves at the speed light in the direction of propagation of the relativistic stream). All simulations with $n_b/n_0=1$ except simulation \emph{$nb = 0.1$}. All simulations with $\gamma =10$, except simulation \emph{$u=5$}. All simulations with steep profile of the leading edge of the relativistic plasma shell, except simulation \emph{smooth} (profile rises from 0 to 1 in 100 $\lambda_e$. Simulation box has the same features of the simulation box in configuration I.}
\label{fig:13}
\end{figure}

Simulations with a moving window (configuration II) show the same behavior. In Fig. \ref{fig:13} the ratio between the energy in the magnetic field and the initial kinetic energy of the particles in the simulation box is plotted as a function of time. Sub-equipartition levels of the magnetic field are maintained for very long times inside the simulation box ($t \simeq 2000 \tau_e$). This level of magnetic field is maintained by the beam erosion and the relativistic plasma shell slowdown. Even though it is not possible to follow the field evolution behind the simulation box, the runs in configuration I indicate that the magnetic field will be maintained as long as plasma is being injected behind the region of interest, and that it will decay slowly on a the 1000s' $\tau_e$ time scale. Furthermore, at $t \simeq 2000 \tau_e$, the relativistic plasma shell inside the simulation box already lost more than $10\%$ 
of its total initial kinetic energy to the fields and the background plasma, thus indicating significant energy conversion efficiency from  the relativistic outflow to the fields and to the background plasma. 

\begin{figure}[hbt]
\epsfig{file=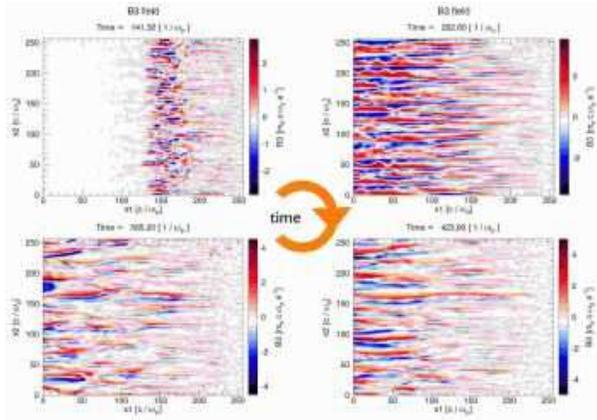,width=8 cm}
\caption{Evolution and structure of the B-field due to the Weibel instability in a moving window simulation box, for the simulation \emph{smooth}.}
\label{fig:14}
\end{figure}

A significant fraction of the field is transferred to the generated B-field, which is also dragged out of the simulation box, as illustrated in Figure \ref{fig:14}, which reproduces the physics already observed in configuration I: a first stage, where the electrons dominate the dynamics and where small scale filaments merge, is followed by a second stage where the ion filaments start to grow. Thus, a strong level of the magnetic field is maintained first by the free energy in the electrons of the shell, and afterwards by the kinetic energy of the relativistic ions. 

The simulations discussed here confirm previous simulations with electron-positron plasmas/ electron-very light proton plasmas, but here we employ very large simulation boxes for very long times, and with more realistic mass ratios. I stress that for a stationary plasma density $n_0 = 10 \, \mathrm{cm^{-3}}$, the simulation box length/width captures $\simeq \, 4 \times 10^7\, \mathrm{km}$, and the simulation run time covers $\simeq \, 25 \times 10^{-3}\mathrm{s}$. Pushing the limits of presently available computing power, it is possible to perform simulations with an order of magnitude increase both in the simulation area and in the simulation time. 

\section{Particle acceleration}

Plasma dynamics at the collisionless scales is also of significant relevance for particle acceleration in relativistic outflows. Not only many plasma mechanisms, associated with wave-particle interactions, can lead to particle acceleration but also plasma instabilities are critical in the initial stages of shock formation, and the subsequent acceleration processes in shocks. Three-dimensional PIC simulations are now very close to be able to address critical questions in the formation of relativistic shocks and particle acceleration in shocks, thus  opening the way to answer how collisionless processes affect particle acceleration, or how particles are accelerated to relativistic energies so that they can can be picked up by the shock.

From the point of view of PIC simulations, particle acceleration studies are more demanding than the study of B-field generation since a very large number of particles is required in order to obtain significant statistics, as well as longer run times to allow for the self-consistent shock formation. The alternative of starting the simulations with a preformed shock structure, while viable for nonrelativistic shocks, it is not feasible for relativistic shocks due to the poorly understood kinetic features of these nonlinear structures, even if a good picture of the macroscopic features of relativistic shocks already exists \cite{blandford1976}.

\begin{figure}[hbt]
\epsfig{file=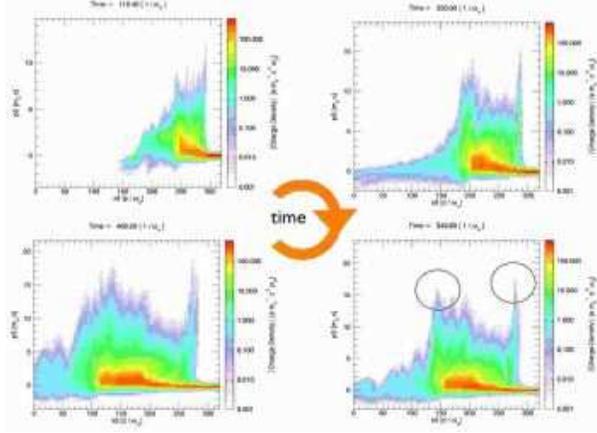,width=8 cm}
\caption{Evolution of the electron $p_3 x_3$ phase space in a moving window simulation box of the collision of electron-positron plasma shells. The simulation box with $(64)^2 \times 1600$ cells, $(12.8)^2\times 320 \left(c/\omega_{pe0}\right)^3$, with the longest direction along the $x_3$ direction, and with 8 particles/cell per species. The relativistic electron-positron shell moves with $\gamma=10$ with the same density as the background electron-positron plasma.}
\label{fig:15}
\end{figure}

In order to understand the possibility of particle acceleration in the early stages of shock formation, we have performed 3D simulations with very long simulations boxes using configuration II (moving window) for the collision of electron-positron shells (Figure \ref{fig:15}). Particle acceleration was clearly observed in the contact regions between the shells, associated with a longitudinal electric field. The electric field is generated by the two stream instability occurring between the relativistic shell and the stationary plasma.

\begin{figure}[hbt]
\epsfig{file=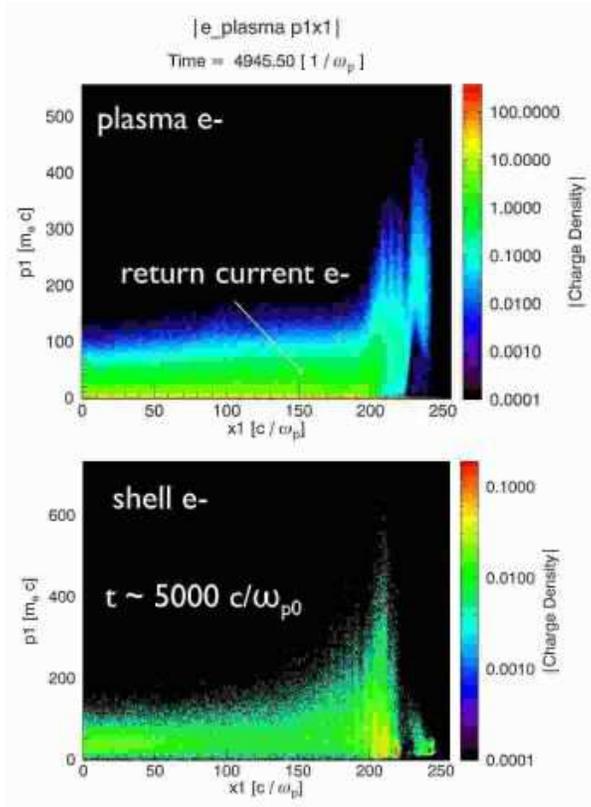,width=8 cm}
\caption{Electron $p_1 x_1$ phase space in a moving window simulation box of the collision of electron-proton plasma shells: in the top frame, stationary background electrons; in the bottom frame, electrons   in the relativistic shell. Results refer to simulation \emph{smooth} discussed in the previous section.}
\label{fig:15a}
\end{figure}

The same scenario occurs for electron-proton shells, as seen in 2D$\|$ simulations. We observe the generation of the longitudinal accelerating structure due to the onset of Buneman instability \cite{davidson} of the relativistic ions on the background plasma. In fact, and due to the electron Weibel instability, a strong erosion of the electrons in the relativistic plasma shell is present. Charge neutrality in the relativistic shell is maintained by the electrons of the background which are picked  up and accelerated by the longitudinal electric field generated in the leading edge of the relativistic shell (Figure \ref{fig:15a}).

\begin{figure}[hbt]
\epsfig{file=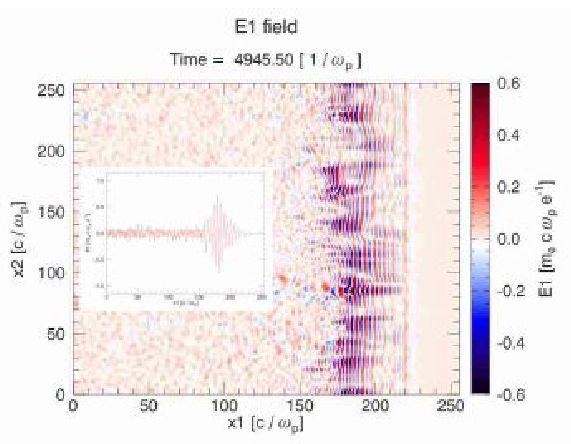,width=8 cm}
\caption{Longitudinal electric field in the leading edge of the relativistic shell. The inset shows the lineout along the center of the box. Results refer to simulation \emph{smooth} discussed in the previous section.}
\label{fig:16}
\end{figure}

The longitudinal electric field structure generated in the leading edge of the relativistic shell is shown in Figure \ref{fig:16}, with the typical wavenumber associated with Buneman instability, $k_\| \simeq \omega_{pe0/c} = 1/\lambda_e$, clearly identifiable in the lineout of the longitudinal electric field (it is important to stress that the growth rate of the Buneman instability scales with $\sqrt{n_b/n_0}\sqrt{m_e/m_i}$, thus involving a very long time scale). 
The maximum electric field $E_\mathrm{max}$ is very close to the nonrelativistic wavebreaking limit \cite{dawson1959}, $E_0 = c m_e \omega_{pe0}/e$, or $E_0 [\mathrm{V/cm}]\simeq 0.96 \sqrt{n_0[\mathrm{cm^{-3}}]}$ (in our normalized units $E_0=1$, and $|E_\mathrm{max}| \approx 0.6$).

With such a strong peak electric field,  electrons can be picked up and reflected by the electric field structure, with a maximum energy given by $\gamma_\mathrm{max} \simeq 4 \gamma^2_\mathrm{shell} |E_\mathrm{max}|$, in excellent agreement with the simulation results, cf. Figures (\ref{fig:15a},\ref{fig:16}). This allows electrons to be strongly accelerated to energies much higher than those reached in the region where filaments merge and cross \cite{hededal2004}. 

These results clearly indicate that even in the early stages of relativistic shock formation, plasma dynamics at the collisionless scale plays an important role contributing to strong electron acceleration. The highly relativistic electrons resulting from acceleration in this region can then be easily injected into relativistic shocks and be further accelerated to very high energies. The accelerating structure observed in the shock precursor region is present either in electron-positron plasmas or in electron-proton plasmas, indicating that, in general, a fraction of the free energy in the plasma shells can be converted, due to wave-particle interactions, into a population of electrons accelerated to energies in the order of $\gamma_{shell}^2$.

\section{Conclusions and the \emph{Open Question}}

I have revised the current advances in the study of the microphysics relevant to relativistic outflows. Particular emphasis was given to the collisionless mechanisms arising in the collision of relativistic shells, and the impact that PIC simulations are having in the study of these scenarios.
 
Sub-equipartition magnetic fields have been identified in multi-dimensional simulations, capable of capturing different aspects of the instabilities responsible for the transfer of free energy from the plasma shells to the fields, for time scales and lengths scales that cover the dynamics of the heavier species.

The acceleration mechanisms associated with collisionless instabilities have also been discussed. Generation of longitudinal electric fields, due to the Buneman instability, close the the nonrelativistic wavebreaking limit have been shown to provide an acceleration site to high energy electrons.

The results presented here summarize the present state-of-the-art kinetic studies of these scenarios. The advent of petascale computing will open new opportunities to probe these relativistic scenarios with PIC simulations well beyond the scales discussed here, thus opening the way to address the challenging problem of \emph{self-consistent "first principles"} simulations of shock formation, and, most important and critical in astrophysical scenarios, particle acceleration in relativistic shocks.


\begin{theacknowledgments}
 I would like to thank Prof. Ricardo Fonseca, Dr. Gianfranco Sorasio, Prof. Warren Mori, Prof. Mikhail Medvedev, Dr. John Tonge, Professor Chuang Ren,  Massimiliano Fiore, Michael Marti, and Michail Tzoufras for the fruitful collaborations that have contributed to the work described here, as well as Dr. Anatoly Spitkovsky for discussions. This work was partially supported by FCT (Portugal).
\end{theacknowledgments}



\bibliographystyle{aipproc}   


\end{document}